\documentstyle[12pt]{article}

\renewcommand{\Large}{\large}

\begin{document}

\def\be{\begin{equation}}
\def\ee{\end{equation}}
\def\ba{\begin{array}{l}}
\def\ea{\end{array}}
\def\bea{\begin{eqnarray}}
\def\eea{\end{eqnarray}}
\def\eq#1{(\ref{#1})}
\def\del{\partial}
\def\A{{\cal A}}
\def\sh{{\rm sinh}}
\def\ch{{\rm cosh}}

\renewcommand\arraystretch{1.5}

\begin{flushright}
TIFR-TH-96/26\\
hep-th/9605234
\end{flushright}
\begin{center}
\vspace{3 ex}
{\Large\bf 
ABSORPTION VS DECAY OF BLACK HOLES IN STRING THEORY AND T-SYMMETRY
}\\
\vspace{10 ex}
Avinash Dhar, Gautam Mandal and Spenta R. Wadia \\
Tata Institute of Fundamental Research \\
Homi Bhabha Road, Bombay 400 005, INDIA \\
\vspace{1 ex}
e-mail: adhar, mandal, wadia@theory.tifr.res.in\\
\vspace{15 ex}
\bf ABSTRACT\\
\end{center}
\vspace{2 ex}
Classically a black hole can absorb but not emit energy. We discuss
how this T-asymmetric property of black holes arises in the recently
proposed (T-symmetric) microscopic models of black holes based on
bound states of D-branes. In these string theory based models, the
nonvanishing classical absorption is made possible essentially by the
exponentially increasing degeneracy of quantum states with mass of the
black hole. The classical limit of the absorption crosssection 
computed in the microscopic model agrees with the result obtained from
a classical analysis of a wave propagating in the background metric 
of the corresponding black hole (upto a numerical factor).
\vfill
\eject

\setcounter{section}{-1}

\section{Introduction} 

Recent rapid developments in string theory have opened up the exciting 
possibility of a microscopic derivation of the physics of black holes.
Based on progress in understanding bound states of D-branes 
\cite{pol9510,wit9510,sen9511,vaf9511,bersadvaf,dou9512},
microscopic models of black holes have been constructed in 4+1 
\cite{strvaf,calmal,hormalstr,bre4vaf,horstr}
and 3+1 \cite{malstr,johkhumye} dimensions.  Perhaps the most
promising feature of these models is that a counting of microscopic
states correctly reproduces black hole degeneracy as required by the
Bekenstein-Hawking formula $S= \frac{1}{4} A$ \cite{hawking}. There
are, of course, many other aspects of black hole physics which one
would like to derive from these simple microscopic models. The very
existence of an event horizon is one such aspect. This implies that
classically a black hole can absorb but not emit energy. Can one
possibly understand this apparent lack of time-reversal
symmetry\footnote{
The fact that a black hole horizon leads to the abovementioned
time-asymmetry has long been recognized in the literature
\cite{hawking2,penrose}. See \cite{penrose}\ for a rather detailed
discussion and review.}  in terms of the proposed microscopic models
which are based on a manifestly time-reversal symmetric microscopic
theory?

A priori it would seem that this question must remain unanswered at
present. This is because while the microscopic models of black holes
as bound states of D-branes have been constructed in the weakly
coupled string theory regime, the semiclassical picture of the 
black hole is expected to be valid in the strong coupling regime
\cite{strvaf}, 
and exploring this regime is at present beyond our technical
abilities. It is, therefore, surprising that the simple microscopic
model of the Hawking emission proposed in \cite{calmal,malsus} yields
an expression for the decay rate that agrees with the standard formula
in all its essential details! Even though we don't quite understand
why this works, we are encouraged enough by this agreement to address
the question about the classical absorption by a black hole as a first
step towards exploring the horizon physics in the microscopic models.
In this paper we will consider the microscopic model of
\cite{calmal,malsus} and show that the time-reversed process of the
Hawking emission leads to absorption by the black hole which is indeed
nonzero in the classical limit, unlike the Hawking emission which
vanishes in this limit. This leads to classical absorption but not
decay by this microscopic model of the black hole.

This paper is organized as follows. In Section 1 we review the model
of \cite{calmal,malsus} for Hawking emission. In Section 2 we consider
the time-reversed process and calculate the absorption coefficient. At
the microscopic level the magnitude of the matrix element leading to
decay is identical to that leading to absorption. Nevertheless, the
classical limit of absorption by the macroscopic black hole is nonzero
while the decay vanishes. In Section 4 we discuss the classical
propagation of a massless scalar particle in the geometry of the black
hole under discussion. We calculate the absorption crosssection and
compare with the result of Section 3. In Section 5 we discuss possible
strong coupling effects. We argue that the essential details of the
result obtained here are expected to survive even in the strong
coupling limit.

\section{Hawking Decay}

In this section we will review the model of Hawking decay of the
4+1-dimensional charged black hole considered in
\cite{calmal,malsus}. We will
use a notation that will be useful later for discussing absorption in
the next section.

In the microscopic model the decay of a black hole is interpreted as
the annihilation of two massless open string excitations on a D-brane,
each with energy $\omega/2$, into a massless closed string quantum of
energy $\omega$.  As discussed in \cite{calmal}, these open string
degrees of freedom live in the 2-dimensional space-time whose space
part is the $S^1$ of radius $R$ along $x^5$ that is common to
D-onebranes (which wrap around the compact coordinate $x^5$ in the
10-dim. space-time) and D-fivebranes (which wrap around the compact
5-dim. space $S^1 \times T^4$ labelled by $x^5, x^6, x^7, x^8$ and
$x^9$). There are $N_B$ species of bosonic (and as many fermionic)
open string levels for each value of momentum in the $x^5$
direction. $N_B = Q_1 Q_5$ for the extremal holes where $Q_1$ is the
number of D-onebranes and $Q_5$ is the number of D-fivebranes. As
discussed in \cite{malsus}, however, there are problems in applying
this naive picture to realistic black holes, the so-called ``fat''
black holes. These problems are resolved in a modified model proposed
in \cite{malsus}, which uses an observation made in \cite{dasmat}. In
this modified model the D-onebranes and D-fivebranes arrange themselves
inside the bound state in such a way that effectively the number of
bosonic species $N_B = 1$, while the effective radius of the $S^1$
along the $x^5$ direction in which these bosons live is $L = R Q_1
Q_5$. Since the charged black holes considered in \cite{calmal} and under
discussion here, especially in sections 3 and 4 where we discuss the
classical limit, are of the ``fat'' type, we will henceforth work
within the effective model of \cite{malsus}.

We may write a low-energy effective action for the interaction of a
closed massless (in the 4+1 noncompact space-time, $\vec x = (x^1,
x^2, x^3, x^4)$ and $t$) string with two massless (in the 2-dim.
space-time $x^5$ and $t$) open strings:
\be
S_{int} \propto g_{st} \int dt \ \int^{2\pi L}_0 dx^5 \del_a \phi (t,
x^5) \del^a \phi (t, x^5) h (t, \vec x = 0), \ \ 
a = (t, x^5)
\label{1}
\ee
Such an interaction can be inferred from the analyses of 
\cite{garmye,haskle,kletho,gubhasklemal}.
In the above, dimensional reduction has been applied to the directions
$x^6, x^7, x^8$ and $x^9$ and hence the fields are independent of
these coordinates. The field $\phi (t, x^5)$ describes a massless open
string excitation\footnote{ 
This field is assumed to be bosonic. In principle one should also
consider fermionic open string excitations. In this work we have
considered emission and absorption of bosonic closed strings
only. Since the contribution of fermionic open string excitations in
this case is much smaller than the contribution of bosonic open string
excitations at the energies of interest to us, we have ignored them.
However, the fermionic excitations do contribute to black hole
degenracy and we have included their contribution to the counting of
microstates.
}  moving along $x^5$. The field $h (t, \vec x)$ is a massless (scalar)
closed string --- its masslessness in the (4+1)-dimensional noncompact
space-time $(\vec x, t)$ implies that it is independent of
$x^5$. Finally, $g_{st}$ is the string coupling in 10-dimensions and
$L = R Q_1 Q_5$ is the effective radius of the $S^1$ along $x^5$.

The normal mode expansion of the field $\phi (t, x^5)$ is 
\be
\phi (t, x^5) = \sum^{+ \infty}_{h = - \infty} \ {1 \over \sqrt{2
\omega L}} \left( a_n \exp[{i nx^5 \over L} - i \omega t] + c.c.\right)
,\  \omega = {|n| \over L} 
\label{2}
\ee

\vspace{3 ex}

\noindent\underbar{\bf Matrix element for decay} 

\vspace{3 ex}

Let us now consider an initial state $| i \rangle$ with total left-moving
(along $x^5$) momentum $\frac{\tilde N_L}{L}$ and total right moving
momentum $\frac{\tilde N_R}{L}$. There are, of course, many states
$| i \rangle$ corresponding to a given choice of $\tilde N_L$ and $\tilde
N_R$. These are characterized microscopically by the set of number $\{
\tilde N_L (n), \ n = 1, 2, \cdots , \infty \}$ and $\{ \tilde N_R
(n), \ n = 1, 2, \cdots , \infty \}$ where $\tilde N_L (n) = a^+_n a_n
\;, \ \tilde N_R (n) = a^+_{-n} a_{-n}\;, \ n = 1, 2, \cdots ,
\infty$. In other words,
\be
| i \rangle = \prod^\infty_{n=1} \left(\tilde N_L (n) ! \; \tilde N_R (n) !
\right)^{- {1 \over 2}} (a^\dagger_n)^{\tilde N_L (n)} 
(a^\dagger_{-n})^{\tilde N_R (n)} | 0 \rangle
\label{3}
\ee
Clearly, $ \tilde N_{L,R} = \sum_{n=1}^\infty n \tilde N_{L, R}(n) $.

It is a matter of simple one-dimensional thermodynamics to compute
the number of microstates $| i \rangle$ corresponding to a given choice of
$\tilde N_L$ and $\tilde N_R$. Including contribution from fermionic 
open string excitations also, we get 
\be
\Omega = e^S \ , \ \  S = 2 \pi \left(\sqrt{\tilde N_L} + \sqrt{\tilde
N_R}\right) .
\label{4a}
\ee
This is related to the expression for degeneracy of nonextremal
states given in \cite{hormalstr,calmal} by the relation \cite{malsus} 
\be
\tilde N_{L, R} = Q_1 Q_5 N_{L, R} .
\label{4b}
\ee

The final state $| f \rangle$ that we are interested in is obtained from the
initial state $| i \rangle$ by the annihilation of a left-moving open string
of momentum $\frac{m_L}{L} = \frac{\omega}{2}$ with a right-moving
open string of momentum $\frac{m_R}{L} = \frac{\omega}{2}$ into a
massless closed string quantum of energy $\omega$. The remaining gas
of open string excitations in the final state $| f \rangle$ is, therefore,
characterized by $\tilde N'_L = \tilde N_L - m_L \ , \ \ \tilde N'_R =
\tilde N_R - m_R $ :
\be
| f \rangle = h^\dagger_\omega  \otimes \prod^\infty_{n=1} 
\left(\tilde N'_L (n) ! \
\tilde N'_R (n) ! \right)^{- {1 \over 2}} (a^\dagger_n)^{\tilde N'_L (n)}
(a^\dagger_{-n})^{\tilde N'_R (n)} | 0 \rangle
\label{5}
\ee
Since we will be interested in closed strong quanta with zero momentum
parallel to the branes, for us $m_L = m_R = m$. 

As in  \eq{4a},\ \eq{4b}\ we can easily compute the number of
microstates $| f \rangle$ corresponding to a given choice of $\tilde
N'_L$ and $\tilde N'_R$ :
\be
\Omega' = e^{S'} \ , \ \ S' = 2\pi \left(\sqrt{\tilde N'_L} +
\sqrt{\tilde N'_R} \right) 
\label{6}
\ee

The $S$-matrix element for decay from the initial state $| i \rangle$
to the final state $| f \rangle$, to 1st order in string coupling
$g_{st}$, can be computed using  (1) following standard
perturbation theory rules and is given by
\be
\langle f |S| i \rangle \propto
\frac{g_{st}}{\sqrt{m_L}\sqrt{m_R}\sqrt{\omega V_4}} 
\frac{m_L}{L} \cdot \frac{m_R}{L} \cdot L \delta_{m_L, m_R} \delta
(\omega - \frac{m_L}{L} - \frac{m_R}{L})\, (\tilde N_L (m_L)
\tilde N_R (m_R))^{1/2} 
\label{7}
\ee
where $m_L/L =\omega/2 = m_R/L$ and $V_4$ is the volume of the
4-dimensional noncompact space (box normalization).

Now, recall that the nonextremal black holes under discussion are
characterized by six parameters \cite{hormalstr} which label the
corresponding D-brane bound states. These are denoted $N_1, N_{\bar
1}, N_5, N_{\bar 5}, N_L$ and $N_R$ and stand for respectively the
number of D-onebranes, anti-D-onebranes, D-fivebranes,
anti-D-fivebranes, total left moving momentum and total right moving
momentum. (Note that $Q_1 \equiv N_1 - N_{\bar 1}$ and $Q_5 \equiv N_5
- N_{\bar 5}$.) All microscopic states $| i \rangle$ which have a
common value of these parameters refer to the `same' macroscopic black
hole (`no-hair'\footnote{
For an explicit demonstration that `microscopic' quantum numbers do
not show up for scattering off large black holes, see the analysis of
elementary BPS black holes in heterotic string theory by
\cite{manwad}\ who also showed that hair appears in subleading terms
of $S$ matrix down by inverse powers of mass.  For a more general
analysis of hair in string models of black holes see
\cite{larwil}.}).  Therefore, the microscopic model of the black hole
is a density matrix
\be
\rho = \frac{1}{\Omega} \sum_{\{i\}} | i \rangle \langle i|
\label{8}
\ee
where the sum $\{i\}$ is over all possible distributions $\{\tilde N_L
(n)\}$ and $\{\tilde N_R (n)\}$ keeping $N_L$ and $N_R$ fixed. It is
this formula that leads to the entropy $S = -Tr\rho \ln \rho = -
\sum_{\{i\}} \frac{1}{\Omega} \ln \frac{1}{\Omega} = \ln \Omega$. 

Density matrices like the one in  \eq{8} are not unfamiliar in
particle physics. They arise, e.g. in calculating the decay rate of an
unpolarized particle into unpolarized products. As there, in the
present case also, the total ``unpolarized''transition probability is
given by
\be
P_{\rm decay} (i \rightarrow f) = \frac{1}{\Omega} \sum_{\{i\},\{f\}}
|\langle f|S|i \rangle|^2
\label{9}
\ee
The division by $\Omega$ represents averaging over initial states,
while the final states are simply summed over. The passage to the
decay rate $d\Gamma $ is usual and one gets
\be
d\Gamma  \propto d^4 \vec k G_N m \langle
\tilde N_L (m)\rangle \langle\tilde N_R (m)\rangle
\label{10}
\ee
where $G_N$ is Newton's constant in 4-dimensional noncompact space,
$\omega = \frac{2m}{L}$ is the energy of the emitted massless closed
string, $\vec k$ is its momentum in 4-dimensional space $(|\vec k| =
\omega)$ and $\langle\tilde N_{L,R} (m)\rangle$ is 
the average distribution in the initial state (with fixed total
momenta $\tilde N_L$ and $\tilde N_R$). For large values of $\tilde
N_L$ and $\tilde N_R$ one can compute these average distributions by
approximating the microcanonical ensemble by a canonical ensemble (in
the 1-dimensional thermodynamics that gave rise to \eq{4a} and
\eq{6}). One gets the standard Bose-Einstein distributions
\be
\langle\tilde N_{L,R} (m)\rangle = \left(e^{\beta_{L,R} \omega/2}
- 1 \right)^{-1}
\label{11}
\ee
where
\be
\beta_{L,R} = \pi L / \sqrt{\tilde N_{L,R}} = \frac{\pi L}{\sqrt{Q_1
Q_5 N_{L,R}}}
\label{12}
\ee
and we have used that $m =\omega L/2$. 

\section{Absorption}

Consider now the absorption of a massless closed string quantum by the
black hole. The elementary process here is just the reverse of the
decay process of the previous section. In fact, let us consider the
absorption of a massless quantum of energy $\omega = 2m/L$
by the initial state $| i' \rangle$ labelled by the total momentum
$\tilde N'_L$ and $\tilde N'_R$ (as in the final state $| f \rangle$
of the previous section). The final state $| f' \rangle$ of the
black hole in this case, then, contains an additional left (right)
moving open string mode of momentum $m_L/L =\omega/2$
( $m_R/L = \omega/2$ ) (just like in the initial 
state $| i \rangle$ of the previous section). Thus, in this absorption
process the initial and final states of the previous section just get
interchanged.  Furthermore, it is trivial to see from  (1) (or by
using perturbative unitarity of string theory) that to first order is
$g_{st}$ perturbation theory, $\langle f'|S| i' \rangle = \langle i|S| f
\rangle = - \langle f|S| i \rangle^\ast $. It follows, therefore, that
for the {\it macroscopic} black hole the absorption probability is
\be
P_{\rm abs}(i'\to f') = \frac{1}{\Omega'} \sum_{\{ i'\}, \{f'\}} 
| \langle f' | S | i' \rangle |^2
= \frac{1}{\Omega'} \sum_{\{ i\}, \{f\}} 
| \langle f | S | i \rangle |^2
\label{13}
\ee
The division by $\Omega'$ signifies averaging over the microscopic
initial states $| i' \rangle$ whose degeneracy is the same as that of
the state $| f \rangle$ and is given by $\Omega'$ in  \eq{6}. From
\eq{9} and \eq{13} we see that at the {\it macroscopic} 
level the absorption probability is related to the decay probability
by the equation
\be
P_{\rm abs}(i'\to f') = \frac{\Omega}{\Omega'} P_{\rm decay} (i\to f)
\label{14}
\ee
Thus the absorption probability is {\it larger} than the decay
probability by the factor $\Omega / \Omega'$ (recall that $\Omega$
increases exponentially with the mass of the black hole and that
$\Omega'$ refers to the black hole with mass smaller than that to
which $\Omega$ refers). As we shall see it is this enhancement factor
that is responsible for a nonzero classical absorption by the
black hole. 

The passage from the absorption probability to the absorption
cross-section $\sigma_A$ is usual.  We get
\be
\sigma_A \propto \frac{\Omega}{\Omega'} G_N \frac{\omega L}{2}
\langle \tilde N_L(m) \rangle \langle \tilde N_R(m) \rangle
\label{15}
\ee
where $m = \omega L/2$ and $\langle\tilde N_{L,R} (m)\rangle$
are given in  \eq{11}, \eq{12}. 

\section{Classical Limit}

We would now like to discuss the results of the previous two sections
in the classical limit. This limit is taken by letting the mass of the
black hole become very large, i.e., $M \gg 1$ (in Planck units).
Actually, the classical limit is more subtle in the case of charged
black holes \cite{pre3wil,holwil,krawil}. One also needs to ensure
that the black hole is not too close to extremality. More
quantitatively, for the 5-dimensional charged black holes under
discussion, the appropriate conditions are
\be
M \gg \Delta M \gg \frac{1}{M^2}, \quad M \gg 1
\label{16}
\ee
The second part of the first condition ensures consistency of thermal
description ($\Delta M$ is mass deviation from the extremal limit)
\cite{pre3wil}, 
while the first part ensures that deviations from extremality
are small in the macroscopic sense.

Now, let us assume that the nonextremal charged black holes under
discussion are obtained by perturbing $N_R$ away from its extremal
value of zero. Let us take
the classical limit by scaling $Q_1$, $Q_5 $ and $N (\equiv N_L -
N_R)$ as follows
\be
Q_1 \to \lambda Q_1,\; Q_5 \to \lambda Q_5, \; N \to \lambda N,
\quad \lambda \gg 1
\label{17}
\ee
keeping the ratio $N_R/N_L$ fixed and small.  Such a scaling is
natural for Reissner-Nordstrom black holes which satisfy the condition
$Q_1 R/g_{st} = Q_5 RV/ g_{st} = N/R$.  These scalings have the
following effect on the mass $M$ of the black hole, $\Delta M$ and the
effective radius $L$ of the $S^1$ in the $x^5$ direction:
\be
M \to \lambda M,\; \Delta M \to \lambda \Delta M, \; L \to \lambda^2
L.
\label{18}
\ee
The conditions in \eq{16} are then automatically satisfied for
$\lambda \gg 1$ in the case of absorption, while in the case of decay
they are satisfied at least in the early stages of decay. Thus, a
consistent way of taking the classical limit is to do the scalings
\eq{17}, and let $\lambda$ become large.

Now, under the scalings \eq{17} and \eq{18}, we see from 
\eq{12} that $\beta_{L, R}$ scale as 
\be
\beta_L \to \sqrt \lambda \beta_L,  \qquad  \beta_R \to 
\sqrt \lambda \beta_R
\label{19}
\ee
Note that it follows from \eq{12} that 
$\beta_R$ is always much larger than $\beta_L$ (because of
the condition $N_R \ll N_L$), and remains so under the scalings
\eq{19}.

\vspace{2 ex}

{\bf Vanishing classical decay rate}:

\vspace{2 ex}

We are now ready to discuss the classical limits of \eq{10}\ and
\eq{15}. Let us consider the decay first. Because of  \eq{19}\ the
decay rate peaks at $\omega \sim 1/\beta_R$ in the classical limit.
In  \eq{10} we may, therefore, expand $\langle\tilde N_L
(m)\rangle$ and retain only the first term :
\be
\langle \tilde N_L(m) \rangle \sim \frac{1}{\beta_L \omega}
\label{20}
\ee

Thus, the decay rate becomes \cite{calmal}
\be
d\Gamma \propto d^4 \vec k A_h (e^{\beta_R \omega/2} -1)^{-1}
\label{21}
\ee
where $A_h \sim G_N \sqrt{Q_1 Q_5 N_L}$ is the area of the horizon of
the black hole. Now, using \eq{17}\ and the second of \eq{19}, we find
that the decay rate vanishes exponentially in the classical limit:
\be
d\Gamma \sim \lambda^{3/2} e^{-\sqrt \lambda}
\label{21a}
\ee
where in the last equation we have displayed only the
$\lambda$-dependence.

\vspace{2 ex}

{\bf Nonvanishing classical absorption}:

\vspace{2 ex}

To see what happens to the absorption crosssection, \eq{15}, in this
limit, we also need to compute the enhancement factor
$\frac{\Omega}{\Omega'}$. Using that $\tilde N'_{L,R} = \tilde N_{L,R}
- m$ and $\omega =
\frac{2m}{L}$, we get 
\be
\frac{\Omega}{\Omega'} = e^{[\frac{\omega}{2} \beta_L + o(\omega^2)]}
e^{[\frac{\omega}{2} \beta_R + o(\omega^2)]}
\label{22}
\ee
The coefficient of the $\omega^2$ term in the first exponent
is the derivative of $\beta_L \sim \sqrt{L/M} $ with respect to 
$M$, and under the scalings in \eq{18}\ vanishes as $\lambda^{-1/2}$
for large $\lambda$. Similarly the corresponding coefficient in 
the second exponent involves the derivative of $\beta_R
\sim \sqrt{L/\Delta M}$ with respect to $\Delta M$, which also
vanishes as $\lambda^{-1/2}$ for large $\lambda$. The coefficients of
higher powers of $\omega$ in both the exponents vanish even faster as
$\lambda$ becomes large.\footnote{
Equation \eq{22} is actually valid under a more general scaling of
$\Delta M$ than that given in \eq{18}, namely $\Delta M \to
\lambda^\alpha \Delta M$ where $1 \ge \alpha > 2/3$.}

Using \eq{22}\ and  \eq{11}\ in  \eq{15}, we get 
\be
\sigma_A \propto G_N \frac{\omega L}{2} 
(1 - e^{-\beta_L \omega/2})^{-1}
(1 - e^{-\beta_R \omega/2})^{-1}
\label{23}
\ee
which is clearly nonvanishing in the classical limit.

\vspace{2 ex}

We will now restrict the above formula to frequencies $\omega$
satisfying $\omega \ll \beta_L^{-1}$. This is done for the following reason.
In the classical calculation of the  absorption crosssection in the next
section, we have restricted ourselves to small values of $\omega$. 
The corrections are order $\omega r_0$, where $r_0 \sim (G_N M)^{-1/2}
\sim \beta_L$ is the radius of the horizon. The corrections are, therefore,
higher order in $g_{st}$. To include this consistently one must, therefore,
also include higher order $g_{st}$ corrections in the microscopic model,
which we have not done here.

Now under the condition $\omega \ll \beta_L^{-1}$, we may expand the first
factor in brackets in \eq{23} in powers of $\omega \beta_L$. Retaining only
the first term, we get 
\be
\sigma_A \propto A_h (1 - e^{-\beta_R \omega/2})^{-1}
\label{24}
\ee
Now we let $\lambda$ become large after doing the appropriate scalings
given in \eq{19}. For any given fixed $\omega \ll \beta_L^{-1}$,
$\beta_R \omega $ will eventually become very large as $\lambda$
becomes large\footnote{
The condition $\omega \ll \beta_L^{-1}$ actually puts a restriction
on how large a value of $\lambda$ can be taken. However this does not
affect our conclusion since the maximum value of $\beta_R \omega$
allowed by this condition, namely $\beta_R/\beta_L$, is very large.}
. Therefore in this limit \eq{24} gives
\be
\sigma_A \propto  A_h
\label{25}
\ee
Thus, the enhancement factor $\Omega / \Omega'$ has ensured that the
absorption coefficient remains nonzero in the classical limit! As we
shall see in the next section, the result we have obtained above in
 \eq{25} from a microscopic calculation matches in all its essential
details with that obtained from a classical calculation of wave
propagation in the appropriate black hole geometry. 

\section{Classical Wave Analysis and Absorption}

In this section we consider classical propagation of a massless field
in the geometry of the 4+1 dimensional black hole. We take the
massless field to be one of the scalar moduli which has a simple
propagation equation \cite{sen}
\be
D_\mu \del^\mu \phi = 0
\label{26}
\ee
Here the metric defining the Laplacian is \cite{hormalstr}
\be
\ba
ds^2 = - f^{-2/3}(r) g(r) dt^2 + f^{1/3}(r) [g(r)]^{-1} dr^2 + f^{1/3} 
r^2 [ d\chi^2 + 
\\
\,~~~~~~~~~~~~~~~~~~~~~~~~
\sin^2 \chi d\theta^2 + \sin^2\chi \sin^2 \theta d\phi^2]
\\
g(r) = (1 - r_0^2/r^2)
\\
f(r) = (1 + \frac{r_0^2 }{r^2} \sh^2 \alpha)
(1 + \frac{r_0^2 }{r^2} \sh^2 \gamma)
(1 + \frac{r_0^2 }{r^2} \sh^2 \sigma)
\\
\ea
\label{27}
\ee
The parameters $r_0, \alpha, \gamma, \sigma$ appearing in the
metric can be related to various parameters of the microscopic model
by the relations
\be
Q_1 = \frac{Vr_0^2}{2 g_{st}} \sh 2 \alpha, \;
Q_5 = \frac{r_0^2}{2 g_{st}} \sh 2 \gamma, \;
N = \frac{R^2 V r_0^2}{2 g_{st}^2} \sh 2 \sigma,
\label{28}
\ee
\be
M = \frac{RV r_0^2}{2 g_{st}^2} ( \ch\, 2\alpha + \ch\, 2 \gamma
+ \ch\, 2 \sigma ).
\label{28a}
\ee
In order to calculate the absorption coefficient from \eq{26}, we will
follow a procedure similar to the one used in \cite{unruh} for the 3+1
dimensional Schwarzschild black hole.

Equation \eq{26} admits the following separation of variables:
\be
\phi(r,t,\chi, \theta, \phi) = e^{-i \omega t} R_{\omega l}(r) Z_l(\chi)
\label{29}
\ee
We will be interested in the low frequency behaviour. It is then
enough for us to concentrate on the $s$ wave. The corresponding radial
function $ R_\omega \equiv R_{\omega l}|_{l=0} $ satisfies the
differential equation
\be
[\frac{g}{r^3}\frac{d}{dr} gr^3\frac{d}{dr} + \omega^2 f] R_\omega = 0
\label{30}
\ee
This equation can be alternatively written, in terms of $\psi_\omega
\equiv r^{3/2} R_\omega$, as
\be
\ba
[ - \frac{d^2}{ d r_*^2 } + V_\omega (r_*) ] \psi_\omega = 0
\\
V_\omega (r_*) \equiv -\omega^2 f + 
\frac{3}{4} \frac{1}{r^2} (1 - \frac{r_0^2}{r^2})
(1 + 3 \frac{r_0^2}{r^2})
\\
\ea
\label{31}
\ee
where $r_\ast \equiv \int dr/(1-r^2_0/r^2) = r + (1/2) r_0 
\ln |(r-r_0)/(r+r_0)|$ is the ``tortoise coordinate''.

\vspace{5 ex}

\noindent\underbar{\bf Solution in the Far Region} ($ r \gg
r_0, (r_0/\omega^2)^{1/3} $)

\vspace{5 ex}

In this region we can keep terms only upto $\frac{1}{r^2}$ in 
 \eq{31}.
The solution, given in terms of Coulomb wave functions, has the
following asymptotic expansions:
\noindent (i) $\omega r \gg 1$ 
\be
R_\omega 
\sim r^{-3/2}(-2i\omega)^{-a} \left( e^{-i \omega r} \, 
\frac{\Gamma (2a)}
{\Gamma(a)} A + e^{i\omega r}[ e^{-i\pi a} \frac{\Gamma(2a) } 
{\Gamma(a)}
A + B ] \right)[1 + o(\omega r)^{-1}]
\label{33}
\ee
\noindent (ii) $\omega r \ll 1$ 
\be
R_\omega \sim r^{a - 3/2} e^{i\omega r} [ (A + B \frac{\Gamma(1 - 2a)}
{\Gamma(1 - a)}) - (-2i\omega r)^{1 - 2a} B \frac{\Gamma(1 - 2a)}
{\Gamma(1 - a)} ](1+o(\omega r))
\label{34}
\ee
Here
\be
a = \frac{1}{2} + \sqrt{ 1 - (\omega r_0)^2 (2 + s_1) },
\; s_1 = \sh^2 \alpha + \sh^2 \gamma + \sh^2 \sigma
\ee

\vspace{5 ex}

\noindent\underbar{\bf Solution in the Near Region} ($r \rightarrow
r_0$)

\vspace{5 ex}

Here \eq{30} reduces to 
\be
(g \frac{d}{dr}g \frac{d}{dr} + \omega^2 f_0)  R_\omega = 0
\Rightarrow  [ (\frac{d}{dr_*})^2 + \omega^2 f_0 ] R_\omega = 0
\label{35}
\ee
where
\be
f_0 = f|_{r= r_0} = \ch^2 \alpha \,\ch^2 \gamma \,\ch^2 \sigma
\label{36}
\ee
The solution to \eq{35}, using the boundary condition that there is
no outgoing exponential at the event horizon, is
\be
R_\omega \sim A_0 \exp [ - i ( \omega\sqrt f_0 r_* + \delta ) ]
\label{37}
\ee
Besides these solutions, it is also easy to derive the following exact
$\omega = 0$ solution (at any $r$)
\be
R_{\omega=0} =  A_1 + \frac{B_1}{2} \ln | 1 - \frac{r_0^2} {r^2}|
\label{38}
\ee
Matching \eq{34} and \eq{37} with the $r_0/r \rightarrow 0$ and $r_0/r
\rightarrow 1$ limits of \eq{38} ({\it cf.} \cite{unruh}) we get the
following relations between various coefficients at low frequency
($\omega r_0 << 1$):
\be
\ba
B = \beta A_0 \\
A = [ 1 - \beta \frac{\Gamma (1 - 2a)} {\Gamma ( 1 -a) } ] A_0 \\
\beta \equiv  2 i (\omega r_0)^3 
\sqrt f_0 \frac{ \Gamma (a) \Gamma (2 - 2a)}
{\Gamma (2a) \Gamma(1 - 2a)} 
\ea
\label{39}
\ee
Choosing A to be such that the coefficient of $\frac{e^{i\omega
r}}{r^{3/2}}$ in \eq{33}\ is 1, we get
\be
R_\omega \simeq 
\frac{e^{-i\omega r}}{r^{3/2}} + {{\cal R}} 
\frac{e^{i\omega r}}{r^{3/2}}
\label{40}
\ee
The absorption coefficient is, therefore, to leading order
in $\omega r_0$
\be
| {{\cal A}} |^2 \equiv 1 - |{{ \cal R}} |^2 = \frac{\pi}{2} 
(\omega r_0)^3 \sqrt f_0 = \frac{1}{4 \pi} \omega^3 A_h
\label{41}
\ee
where we have used \eq{36} and the expression for the area of the
event horizon
\be
A_h = 2 \pi^2 r_0^3\, \ch\, \alpha \,\ch \,\gamma \,\ch \,\sigma
\label{42}
\ee

It is easy to show that the $s$ wave absorption crosssection
$\sigma_A$ is related to the absorption coefficient by 
\be
\sigma_A = \frac{4 \pi  }{\omega^3}  |{{\cal A}}|^2
\label{43}
\ee
which in this case, therefore, is
\be
\sigma_A =  A_h 
\label{44}
\ee
as claimed in the previous section. Like \eq{41}, \eq{44}\ is also 
calculated to the leading order in $\omega r_0$.

\section{Concluding Remarks}

In summary, in this letter we have computed the absorption
crosssection of massless quanta by a near extremal 4+1 dimensional
charged black hole within the context of the string theory based
microscopic model proposed in \cite{strvaf,calmal,malsus}. The authors
of \cite{calmal} have correctly reproduced the Hawking radiation
formula for a near extremal black hole (modulo a numerical
coefficient). Our microscopic computation of the absorption
crosssection agrees (modulo a numerical coefficient) with the
classical calculation from the analysis of a massless wave
proapagating in the background metric of the appropriate black hole.

The basic reason why we get a nonzero absorption crosssection is the
presence of the enhancement factor $\Omega / \Omega'$ in this
calculation relative to the Hawking decay, which vanishes in the
classical limit. The factor $\Omega/\Omega'$ depends only on the
counting of the microscopic quantum states of a near extremal black
hole and does not depend on the details of the matrix element
calculation. This is just like the factors $\langle \tilde N_L(m)
\rangle $ and $\langle \tilde N_R(m) \rangle$, 
which also depend only on the counting of states and in the decay
calculation and give rise to the universal black body nature of the
Hawking decay formula. The precise cancellation of these factors with
the enhancement factor $\Omega/ \Omega'$, in the classical limit, is
then what gives a nonzero result for the classical absorption
crosssection, as opposed to the Hawking decay, which vanishes in the
classical limit.

We believe that it is reasonable to expect that the above feature of
our calculation will not be modified by taking strong coupling effects
into account, at least for near extremal black holes. On the other
hand, one may, a priori, not have expected to get a detailed agreement
of the clasical limit of the microscopic calculation with classical
absorption crosssection calculation. That the former agrees with the
latter in all its essential details is, therefore, a surprise. The
magic here is the same as the one that gives the Hawking decay
coefficient proportional to the area of the horizon in the calculation
of \cite{calmal}. This is because the magnitude of the microscopic
matrix element that is responsible for absorption is the same as the
one that gives the decay, at this order of string coupling. It is
possible that with a better understanding of the microscopic models of
black holes we might understand why certain physical situations are
insensitive to the strong coupling effects of the ``dense horizon
soup'' \cite{malsus}.  In this context, it would be very interesting
to compute the numerical coefficient in front of the decay rate in
\eq{21}\ and the absorption coefficient in \eq{25}\ and to see whether
they agree with their expected values.

One of the essential features of the existence of a horizon is that
classically it acts as a one way valve for particles and energy. It
seems to us that the microscopic models which incorporate this feature
must ``know'' about the existence of a horizon in the strong coupling
regime. There are, of course, many other aspects of the physics of
horizon that need to be explored. Hopefully, further study will
provide a better and more detailed understanding of this and other
aspects of black hole physics within the context of the microscopic
models.

\vspace{5ex}

\noindent{Acknowledgement}: 
One of us (SW) would like to acknowledge the Japan Society for the
Promotion of Science (JSPS) for a fellowship and Professors Yoneya,
Kawai and Ninomiya for their excellent hospitality at the University
of Tokyo (Komaba), KEK and the Yukawa Institute for Theoretical
Physics, Kyoto University, respectively. We would like to thank
S.R. Das for a discussion and for pointing out a numerical error in
equation (46) in an earlier version of the paper. We would also like
to thank T.P. Singh for pointing out reference \cite{penrose}.

\def\no#1{{\tt hep-th #1 }}
 
\end{document}